# Parameter estimation of beta-geometric model with application to human fecundability data

B. P. Singh[1], P. S. Pudir[2] and [3]Sonam Maheshwari[3]

[1]Faculty of Commerce, DST-CIMS, Banaras Hindu University, Varanasi

[2]Department of Statistics, Univ. of Allahabad, Allahabad& DST-CIMS, BHU, Varanasi

[3]*Department of Statistics, Banaras Hindu University, Varanasi,

*Corr Author: maheshwarisonam2@gmail.com

**Abstract**

The present study deals with the estimation of the mean value of fecundability by fitting a theoretical distribution from the observed month of first conception of the married women who did not use any contraceptive method before their first conception. It is assumed that fecundability is fixed for a given couple, but across couples it varies according to a specified distribution. Under the classical approach, methods of moment and maximum likelihood is used while for Bayesian approach we use the above two estimates as prior for fecundability parameter. A real data analysis from the third National Family Health Survey (NFHS-III) is analyzed as an application of model. Finally, a simulation study is performed to access the performance of the several of methods used in this paper

## 1. Introduction

The size and composition of population is highly related with fertility rate and growth rate of the population and fertility and growth rate of the population is governed and related by the terms fecundity and fecundability. Fecundity is related to the biological capacity to conceive of woman and fecundability is the probability of conception during a given menstrual cycle of those women who did not use any family planning method before their first conception and are sexually active. Fecundability has an opposite relationship to the time interval required to conceive from marriage to first birth interval. This time interval is also known as conception delay, conception interval and conception wait. Conception interval and fecundability are the two important and inter related fertility parameters and these are regarded as the most direct measures of fertility of a population. Thus the concept of fecundability is one of the principal determinates of fertility and in human reproductive behavior.

Although, the theoretical importance of fecundability is beyond question, there are several difficulties in estimating it from direct observation. Fecundability is frequently estimated from the distribution of waiting time to conception. Gini (1924) first considered birth intervals as waiting time problems dependent on fecundability. His work Deals with pregnancies and birth of first order under constant fecundability. However, It is not realistic to assume constant fecundability as it is to be governed by many socio economic and demographic Variables. In fact, there is enough evidence that couples vary in their fecundability. About 30% of sexually active couples achieve pregnancy in their first non contraception cycle, a smaller proportion of the remaining couples achieve pregnancy in the second, and with each additional unsuccessful cycle, the conception rate continues to decline, as the risk sets become further depleted of relatively fecund couples [Weinberg and Gladen (1986)].

The fecundability parameter is assumed to follow certain distribution as fecundability varies from women to women. One can assume one of many continuous distributions for fecundability lies in the parameter space [0, 1]. Even though number of possible (Kotz & Van Dorp, 2004; Johnson, Kemp, & Kotz, 2005) univariate continuous distributions defined on the standard unit interval [0,1] are available as the mixing distribution of the success probability random variable, the Beta distribution denoted as *Beta* (*a*, *b*), where *a* and *b* are the two shape parameters of the Beta distribution, is the most commonly used mixing distribution to model the random variable defined on the standard unit interval [0,1] due to its ability of accommodating wide range of shapes. Thus the beta-geometric (BG) distribution, represented by *Geo* (theta) ^ *Beta* (*a*, *b*), is considered as a very versatile distribution in modeling human fecundability data in literature.

Extensive literatures exist on the study of beta-geometric distribution. The assumption of a beta distribution for the probability of conception was originally proposed by Henry (1957). For the women of constant fecundability; a geometric distribution is used for the waiting time to marriage till conception while for the women of heterogeneous fecundability the resulting distribution for the waiting time till conception is beta geometric. The parameters of this mixed distribution have practical interpretation. According to Sheps (1964), fecundability affects fertility through its relationship with the average time required for a conception to occur and can also be considered as the transition probability for the passage from the susceptible state to pregnancy. In a homogeneous population, fecundability is equal to the reciprocal of its mean conception delay but for heterogeneous populations, the mean fecundability is usually modeled on two parameters [Jain (1969)]. Weinberg and Gladen (1986) considered that the decrease in conception probability over time is a sorting effect in a heterogeneous population, rather than a time effect. They proposed a general model for heterogeneity and studied in detail the particular case where the distribution for the number of cycles to conception has a beta-geometric distribution. Pual (2005) develop tests of heterogeneity in the fecundability data through goodness of fit of the geometric model against the beta-geometric model along with a likelihood ratio statistic and a score test statistic. Islam et al (2005) also made an attempt to compare the two methods of estimation of the mean value of fecundability using the Bangladesh Demography and Health Survey (1999-2000) data.

In lieu of above considerations, the paper is organized as follows. In section 2, we describe the model by assuming that fecundability is fixed for a given couple, but across couples it varies according to beta distribution and hence obtain the beta-geometric model as the unconditional distribution of the conception delay distribution. In section 3.1, we obtain the moment estimate of fecundability parameter. In section 3.1, we obtain the maximum likelihood estimators (MLE) of the unknown parameter of beta-geometric distribution. It is observed that the MLE is not obtained in closed form, so it is not possible to derive the exact distribution of the MLE. Therefore, we propose to use the asymptotic distribution of the MLE to construct the approximate confidence interval. Further, by assuming beta prior of the fecundability parameter, Bayes estimate is obtained in section 3.3. It is observed that the posterior distribution of fecundability parameter also becomes beta-geometric. In Section 4, demonstrate the application of the beta geometric distribution to the data obtained from the National Family Health Survey (NFHS)-3. In section 5, a simulation study is also carried out to check the performance of classical and Bayesian methods of estimation.

## 2. The Model

After a couple decides to have a child, the number of months elapsed before the time of conception is denoted by X. If the fecundability at each month, $\theta$ stays constant over time for a given couple then X has a geometric distribution

$$P(X = x | \theta) = \theta(1-\theta)^x; 0 \leq \theta \leq 1; x = 0, 1, 2, 3... \quad (1)$$

$E(x) = \frac{(1-\theta)}{\theta}$ and mean fecundability is $\theta = \frac{1}{E(x)+1}$.

This is known as the conditional distribution of conception delay. Now if $\theta$ varies among couples according to beta distribution, then $\theta$ has the following density function

$$f(\theta) = \frac{1}{B(\alpha,\beta)} \theta^{\alpha-1}(1-\theta)^{\beta-1}; \alpha, \beta > 0$$

Where $B(\alpha,\beta) = \frac{\overline{\alpha+\beta}}{\overline{\alpha}\,\overline{\beta}}$ is the beta function, $\overline{\alpha}$ is the gamma function defined as

$$\overline{\alpha} = \int_0^\infty x^{\alpha-1} e^{-x} dx$$

and $(\alpha,\beta)$ are two unknown non-negative parameters. The mean and the variance of beta random variable θ are $\mu = \frac{\alpha}{\alpha+\beta}$ and $\sigma^2 = \frac{\alpha\beta}{(\alpha+\beta)^2(\alpha+\beta+1)}$ respectively.

And the unconditional distribution of the conception delay X is given by

$$P(X = x) = g(x) = \int_0^1 f(x,\theta) d\theta = \int_0^1 P(X = x | \theta) f(\theta) d\theta = \frac{B(\alpha+1, x+\beta)}{B(\alpha,\beta)} \quad (2)$$

This distribution is known as beta-geometric distribution. In the human reproduction literature, P(X=x) is the probability that conception occurs at x for a randomly selected couple. Weinberg and Gladen (1986) written the beta-geometric distribution in terms of the parameter $\pi = \alpha/(\alpha+\beta)$ and $\theta = 1/(\alpha+\beta)$, where p is interpreted as the mean parameter and θ as the shape parameter, and is given by

$$P(Y = y \mid n) = \frac{\pi \prod_{\tau=0}^{y-2} \{(1-\pi) + \tau\theta\}}{\prod_{\tau=0}^{y-1} \{1 + \pi\theta\}}$$

The mean and variance for the above defined form of beta-geometric model is $\frac{1-\theta}{\pi-\theta}$ and variance $\frac{\pi(1-\pi)(1-\theta)}{(\pi-\theta)^2(\pi-2\theta)}$ respectively.

## 3. Estimation of Fecundability Parameter:

In this section, we obtain the estimate of fecundability parameter by using the following three methods of estimation:

### 3.1 Moment Estimation:

Let us suppose that $m_1$ and $m_2$ are the two observed first and second raw moment of the months required for women to conceive for the first time after their marriage. The corresponding population moment of X about origin, conditional on $\theta$, as given by the simple geometric distribution are

$$E(x) = \sum_{x=0}^{\infty} x\theta(1-\theta)^x = \frac{(1-\theta)}{\theta} \text{ and } E(x^2) = \sum_{x=0}^{\infty} x^2\theta(1-\theta)^x = \frac{(1-\theta)}{\theta} + \frac{2(1-\theta)^2}{\theta^2}$$

To obtain the unconditional moment of X, we have to put the value of

$$E\left(\frac{(1-\theta)^r}{\theta^r}\right) = \int_\theta \frac{(1-\theta)^r}{\theta^r} p(\theta) d\theta = \frac{\beta^{[r]}}{(\alpha-1)^{(r)}} \quad ; where \begin{cases} x^{[r]} = x(x+1)(x+2)\ldots(x+r-1) \\ x^{(r)} = x(x-1)(x-2)\ldots(x-r+1) \end{cases}$$

to get

$$\mu_1' = \frac{\beta}{\alpha-1}; \qquad \mu_2' = \frac{\beta(2\beta+\alpha)}{(\alpha-1)(\alpha-2)}$$

Now, equation $m_1$ with $\mu_1'$ and $m_2$ with $\mu_2'$ we get

$$\hat{\alpha} = \frac{2(m_2 - m_1^2)}{m_2 - m_1 - 2m_1^2}; \qquad \hat{\beta} = m_1(\hat{\alpha} - 1) \tag{3}$$

From equation (3), we can easily obtain the moment estimate of $\alpha$ and $\beta$, hence moments estimate of the fecundability parameter $\theta$. As the sample becomes very large, the asymptotic variance-covariance matrix can be obtained similarly those obtained by Islam et al (2005) using the method described by Rao (1952).

### 3.2 Maximum Likelihood Estimation:

Suppose that data are available on N individuals as $x_i; i = 1, 2, \ldots, n$. The likelihood function for data based on beta geometric distribution is given as

$$L = \prod_{i=1}^{n} \frac{B(\alpha+1, x_i + \beta)}{B(\alpha, \beta)} \tag{4}$$

And the corresponding log-likelihood $L(\Theta); \Theta = (\alpha, \beta)$ is given as

$$\log L = L(\Theta) = \sum_i \log B(\alpha+1, x_i + \beta) - n \log B(\alpha, \beta) \tag{5}$$

The score function $U(\Theta)$ is defined as the gradient of $L(\Theta)$, derived by taking the partial derivatives of $L(\Theta)$ with respect to $\alpha$ and $\beta$. The components of the score function $U(\Theta) = (U_\alpha(\Theta), U_\beta(\Theta))^T$ are given below

$$U_\alpha(\Theta) = \frac{\partial L(\Theta)}{\partial \alpha} = N\psi(\alpha+1) + N\psi(\alpha+\beta) - \sum_i \psi(x_i + \alpha + \beta + 1) - N\psi(\alpha) \tag{6}$$

$$U_\beta(\Theta) = \frac{\partial L(\Theta)}{\partial \beta} = \sum_i \psi(x_i + \beta) + N\psi(\alpha+\beta) - \sum_i \psi(x_i + \alpha + \beta + 1) + N\psi(\beta) \tag{7}$$

The maximum likelihood estimates $\alpha$ and $\beta$ can be obtained either by directly maximizing the above log likelihood function with respect to $\Theta$ or by solving the two simultaneous equations obtained by equating $U(\Theta) = 0$. From equations (6) and (7) we can see that the MLEs of $\Theta = (\alpha, \beta)$ cannot be obtained in closed form. Therefore, we need some numerical iterative procedures such as Newton-Raphson method. One can also use the vglm function of R-Environment to obtain the MLE of $\Theta = (\alpha, \beta)$. Using the invariance property of MLEs, one can easily obtain the MLEs of fecundability parameter $\theta$. Further, the asymptotic sampling distribution of $(\hat{\alpha} - \alpha, \hat{\beta} - \beta)'$ is $N_2(0, \Delta^{-1})$, where $\Delta$ is the Fisher's information matrix consisting of the following elements:

$$\Delta_{11} = -\left.\frac{\partial^2 \log L}{\partial \alpha^2}\right|_{\alpha=\hat{\alpha}} \quad \Delta_{22} = -\left.\frac{\partial^2 \log L}{\partial \beta^2}\right|_{\beta=\hat{\beta}} \text{ and } \Delta_{12} = \Delta_{21} = -\left.\frac{\partial^2 \log L}{\partial \alpha \partial \beta}\right|_{\alpha=\hat{\alpha},\beta=\hat{\beta}}$$

The asymptotic $(1-\gamma)\times 100\%$ confidence intervals (C.I.) for $\Theta=(\alpha,\beta)$ is $\hat{\Theta} \pm z_{\gamma/2}\sqrt{V(\hat{\Theta})}$. Here $V(\hat{\Theta})$ is the variance of $\hat{\Theta}$ obtained from $\Delta$ and $z_{\gamma/2}$ is the upper $100\times(\gamma/2)^{th}$ percentile of a standard normal distribution. The respective asymptotic distribution of fecundability parameter is $N(0, \theta'\Delta^{-1}\theta)$ where, $\theta' = \left(\frac{\partial \theta}{\partial \alpha}, \frac{\partial \theta}{\partial \beta}\right)$.

### 3.3 Bayesian Estimation:

In many practical situations, it is observed that the behavior of the parameters representing the various model characteristics cannot be treated as fixed constant throughout the life period. In the introductory section, we have already discussed that the fecundability parameter should not be assumed constant as it is governed by various socio-economic and demographic variables. Therefore, it would be reasonable to assume the parameters involved in the model as random variables. Keeping in mind this fact, we have also conducted a Bayesian study by assuming the following beta prior for fecundability parameter $\theta$

$$h(\theta) = \frac{1}{\beta(\mu,\nu)} \theta^{\mu-1}(1-\theta)^{\nu-1} \tag{8}$$

Here also, the reason of choosing the beta prior for the fecundability parameter is straightforward as it is the most commonly used mixing distribution to model the random variable defined on the standard unit interval [0,1] due to its ability of accommodating wide range of shapes.

Here the hyper parameters $\mu$ and $\nu$ are assumed to be known real numbers. Based on the above prior assumption, the joint density function of the sample observations and $\theta$ becomes

$$L(\underline{x},\theta) = \theta^n (1-\theta)^{\sum_i x_i} \frac{1}{\beta(\mu,\nu)} \theta^{\mu-1}(1-\theta)^{\nu-1} \tag{9}$$

Thus, the posterior density function of $\theta$, given the data is given by

$$\pi(\underline{x}|\underline{x}) = \frac{L(\underline{x}|\theta)h(\theta|\mu,\nu)}{\int_0^1 L(\underline{x}|\theta)h(\theta|\mu,\nu)d\theta} \tag{9}$$

Putting the expression of equation (8) and (9) in equation (10), we get the posterior density of θ

$$\pi(\theta|\underline{x}) = \frac{1}{\beta\left(n+\mu, \sum_i x_i + \nu - n\right)} \theta^{n+\mu-1}(1-\theta)^{\sum_i x_i + \nu - n - 1} \tag{11}$$

It is to be noted that one more plus point of choosing beta prior for fecundability parameter is that it is conjugate prior i.e. the posterior distribution also comes beta. For the squared error loss, the Bayes estimator is the posterior mean and the mean fecundability is

$$\theta^* = \frac{n+\mu}{\sum_i x_i + \nu - n}$$

Without loss of generality, one can assume the value of $\mu$ and $\nu$ obtained by method of moment and maximum likelihood.

## 4. Application of the Model

Here we demonstrate the application of the beta geometric distribution to the data obtained from the National Family Health Survey (NFHS)-3. NFHS-3 was conducted under the stewardship of the Ministry of Health and Family Welfare (MOHFW), Government of India, and is the result of the collaborative efforts of a large number of organizations. The International Institute for Population Sciences (IIPS), Mumbai, was designated by MOHFW as the nodal agency for the project. Funding for NFHS-3 was provided
by the United States Agency for International Development (USAID), DFID, the Bill and Melinda Gates Foundation, UNICEF, UNFPA, and MOHFW. Macro International, USA, provided technical assistance at all stages of the NFHS-3 project.

    A uniform sample design was adopted in NFHS-III. In this Study, only women who had ever been married in age group 15-49 were used. In order to estimate the fecundability for women, we have extracted 3767 women out of 12183 women who have had at least one recognizable conception (regardless of outcome). We have excluded women who were pregnant before marriage. Since our study is based on birth history data, we exclude those

conceptions of women occurring more than 5 years preceding the survey to avoid memory lapse of the respondents. Finally, we have also excluded those women who did not conceive during their first 15 years or 180 months of marriage, because women who fail to conceive within 15 years of their marriage are considered to be the primarily sterile.

So from the above data, we have the following:

$$\sum_x x = 78112; n = 3767$$

Here, $E[x] = 20.73586$ and hence the mean fecundability is $20.73586$.

Now the estimated values of the parameters involved in the model obtained by using the different method of estimation are given as follows:

Method of Moment: $\hat{\alpha} = 20.4669; \hat{\beta} = 403.663 \Rightarrow \hat{\theta} = 0.04825621$

Method of MLE: $\hat{\alpha} = 20.94735; \hat{\beta} = 413.6093 \Rightarrow \hat{\theta} = 0.04820394$

Bayes Estimation: $\hat{\theta} = 0.0460185$ (Using Moment Estimates Prior)

$\hat{\theta} = 0.04601841$ (Using ML Estimate as Prior)

All the methods of estimation are near about the theoretical value of fecundability. Although, estimated mean value of fecundability for Bayes estimate (Using ML estimate as prior) is much closer to the true value, and hence, we can say that Bayes procedure is best for the above data set. Bayes estimate of fecundability using ML estimate as prior is near to true value than those obtained by using moment estimate as prior. Among method of moment and maximum likelihood, MLE perform better than MME in terms of true value.

## 5. A Simulation Study:

Here, we assess the performance of the moment estimate, maximum-likelihood estimate and Bayes estimate of mean time to fecundability with respect to sample size *n*. The measures that are employed for the comparative study of estimation methods for the model parameters are-

- MMEs, MLEs and Bayes estimates.
- Biases and Mean Square Errors (MSEs)

For each of the following options, we simulated six sets of data with samples of sizes 100, 200, 400, 500, 750 and 1,000 respectively, and based on each set of data we computed the above mentioned measures.

- $\alpha = 4, \beta = 36 \Rightarrow \theta = 0.1$
- $\alpha = 4, \beta = 12 \Rightarrow \theta = 0.25$
- $\alpha = 4, \beta = 4 \Rightarrow \theta = 0.5$
- $\alpha = 4, \beta = 2 \Rightarrow \theta = 0.67$

The above assessment is based on following algorithm:

(1) Generate 5,000 samples of size *n* from beta-geometric distribution. The inversion method cannot be used to generate random sample from the beta-geometric distribution as the cumulative distribution function is an incomplete beta (i.e. incomplete gamma) function. To overcome this problem, we use VGAM package of R-environment.

(2) Compute the moment estimate, maximum likelihood estimate and Bayes estimate for the 5, 000 samples, say $\hat{\theta}_i$ for i=1, 2,......., 5000.

(3) Compute the Average Estimates (AE), biases and mean-squared errors given by

$$Bias(\theta) = \frac{1}{5000} \sum_{i=1}^{5000} \left(\hat{\theta}_i - \theta\right)$$

And

$$MSE(\theta) = \frac{1}{5000} \sum_{i=1}^{5000} \left(\hat{\theta}_i - \theta\right)^2$$

(4) We repeat these steps for *n* =100,200 . . . , 1000 with $\theta$ = 0.67, 0.5, 0.25 and 0.10 hence computing AE, bias and *MSE* for *n* = 100, 200, . . . , 1000.

Form the figure and tables 1-4, the following observation can be made

- For all the four choices of the fecundability parameter $\theta$, The magnitude of the Bias and MSE decreases as the sample size *n* increases thereby leading to increased precision in the estimation of the fecundability parameter.
- Though all the considered methods of estimation are precisely estimating the parameters, Bayes estimate of the fecundability with MME prior are consistently performing better than the other methods studied. This is true for all the four options of the fecundability parameter $\theta$.

- Bayes estimate of the fecundability with MME prior is more or less coincide with the Bayes estimate but the precision in estimation is more in Bayesian estimation with MME prior both in terms of Bias and the MSE.
- The biases are negative for the method of moment and maximum likelihood while it is positive for Bayesian method of estimation.
- The biases appear largest for $\theta = 0.5$ while it decreases for other values.

Table1: Average Estimate of $\theta$ with their Bias and MSE for alpha=4; beta=36; theta=0.1 and varying sample size n:

| n | Average Estimate of $\theta$ | | | | Bias of $\theta$ | | | | MSE of $\theta$ | | | |
|---|---|---|---|---|---|---|---|---|---|---|---|---|
| | MME | MLE | Bayes 1 | Bayes 2 | MME | MLE | Bayes 1 | Bayes 2 | MME | MLE | Bayes 1 | Bayes 2 |
| 100 | 0.09467 | 0.10029 | 0.10028 | 0.10013 | -0.00533 | 0.00029 | 0.00028 | 0.00013 | 0.00019 | 0.00022 | 0.00009 | 0.00009 |
| 200 | 0.09606 | 0.09995 | 0.09995 | 0.10012 | -0.00394 | -0.00005 | -0.00005 | 0.00012 | 0.00011 | 0.00011 | 0.00004 | 0.00004 |
| 300 | 0.09683 | 0.09982 | 0.10009 | 0.10015 | -0.00317 | -0.00018 | 0.00009 | 0.00015 | 0.00008 | 0.00007 | 0.00003 | 0.00003 |
| 400 | 0.09734 | 0.09988 | 0.10018 | 0.10022 | -0.00266 | -0.00012 | 0.00018 | 0.00022 | 0.00006 | 0.00005 | 0.00002 | 0.00002 |
| 500 | 0.09771 | 0.09998 | 0.10019 | 0.10021 | -0.00229 | -0.00002 | 0.00019 | 0.00021 | 0.00005 | 0.00004 | 0.00002 | 0.00002 |
| 750 | 0.09824 | 0.10000 | 0.10010 | 0.10012 | -0.00176 | 0.00000 | 0.00010 | 0.00012 | 0.00004 | 0.00003 | 0.00001 | 0.00001 |
| 1000 | 0.09856 | 0.09994 | 0.10007 | 0.10007 | -0.00144 | -0.00006 | 0.00007 | 0.00007 | 0.00003 | 0.00002 | 0.00001 | 0.00001 |

Table2: Average Estimate of $\theta$ with their Bias and MSE for alpha=4; beta=12; theta=0.25 and varying sample size n:

| n | Average Estimate of $\theta$ | | | | Bias of $\theta$ | | | | MSE of $\theta$ | | | |
|---|---|---|---|---|---|---|---|---|---|---|---|---|
| | MME | MLE | Bayes 1 | Bayes 2 | MME | MLE | Bayes 1 | Bayes 2 | MME | MLE | Bayes 1 | Bayes 2 |
| 100 | 0.24198 | 0.24972 | 0.25079 | 0.25056 | -0.00830 | -0.00023 | 0.00061 | 0.00059 | 0.00067 | 0.00058 | 0.00032 | 0.00032 |
| 200 | 0.24210 | 0.24981 | 0.25049 | 0.25056 | -0.00790 | -0.00019 | 0.00049 | 0.00056 | 0.00050 | 0.00050 | 0.00024 | 0.00023 |
| 300 | 0.24332 | 0.24967 | 0.25039 | 0.25053 | -0.00668 | -0.00033 | 0.00039 | 0.00053 | 0.00036 | 0.00033 | 0.00015 | 0.00015 |
| 400 | 0.24441 | 0.24991 | 0.25020 | 0.25028 | -0.00559 | -0.00009 | 0.00020 | 0.00028 | 0.00028 | 0.00024 | 0.00012 | 0.00012 |
| 500 | 0.24512 | 0.24997 | 0.25028 | 0.25034 | -0.00488 | -0.00003 | 0.00028 | 0.00034 | 0.00024 | 0.00019 | 0.00009 | 0.00009 |
| 750 | 0.24620 | 0.24984 | 0.25019 | 0.25022 | -0.00380 | -0.00016 | 0.00019 | 0.00022 | 0.00019 | 0.00013 | 0.00006 | 0.00006 |
| 1000 | 0.24684 | 0.24981 | 0.25024 | 0.25025 | -0.00316 | -0.00019 | 0.00024 | 0.00025 | 0.00015 | 0.00010 | 0.00005 | 0.00005 |

| Table 3: Average Estimate of $\theta$ with their Bias and MSE for alpha=4; beta=4; theta=0.5 and varying sample size n: | | | | | | | | | | | | |
|---|---|---|---|---|---|---|---|---|---|---|---|---|
| n | Average Estimate of $\theta$ | | | | Bias of $\theta$ | | | | MSE of $\theta$ | | | |
| | MME | MLE | Bayes 1 | Bayes 2 | MME | MLE | Bayes 1 | Bayes 2 | MME | MLE | Bayes 1 | Bayes 2 |
| 100 | 0.47981 | 0.48717 | 0.49209 | 0.49102 | -0.01257 | -0.00182 | 0.00108 | 0.00135 | 0.00140 | 0.00127 | 0.00091 | 0.00092 |
| 200 | 0.48801 | 0.49859 | 0.50073 | 0.50108 | -0.01199 | -0.00141 | 0.00073 | 0.00108 | 0.00111 | 0.00101 | 0.00061 | 0.00059 |
| 300 | 0.49086 | 0.49931 | 0.50032 | 0.50043 | -0.00914 | -0.00071 | 0.00032 | 0.00043 | 0.00078 | 0.00068 | 0.00040 | 0.00041 |
| 400 | 0.49271 | 0.50009 | 0.50064 | 0.50074 | -0.00729 | -0.00029 | 0.00064 | 0.00074 | 0.00059 | 0.00052 | 0.00031 | 0.00031 |
| 500 | 0.49336 | 0.49977 | 0.50011 | 0.50018 | -0.00664 | -0.00022 | 0.00011 | 0.00018 | 0.00052 | 0.00042 | 0.00025 | 0.00025 |
| 750 | 0.49425 | 0.49948 | 0.50025 | 0.50029 | -0.00575 | -0.00018 | 0.00025 | 0.00029 | 0.00039 | 0.00028 | 0.00017 | 0.00017 |
| 1000 | 0.49544 | 0.49986 | 0.50037 | 0.50039 | -0.00456 | -0.00014 | 0.00037 | 0.00039 | 0.00029 | 0.00020 | 0.00013 | 0.00013 |

| Table 4: Average Estimate of $\theta$ with their Bias and MSE for alpha=4; beta=2; theta=0.67 and varying sample size n: | | | | | | | | | | | | |
|---|---|---|---|---|---|---|---|---|---|---|---|---|
| n | Average Estimate of $\theta$ | | | | Bias of $\theta$ | | | | MSE of $\theta$ | | | |
| | MME | MLE | Bayes 1 | Bayes 2 | MME | MLE | Bayes 1 | Bayes 2 | MME | MLE | Bayes 1 | Bayes 2 |
| 100 | 0.65872 | 0.66612 | 0.66924 | 0.67869 | -0.01379 | -0.00120 | 0.00099 | 0.00078 | 0.00140 | 0.00129 | 0.00096 | 0.00097 |
| 200 | 0.65557 | 0.66598 | 0.66731 | 0.66736 | -0.01110 | -0.00092 | 0.00088 | 0.00069 | 0.00115 | 0.00103 | 0.00077 | 0.00071 |
| 300 | 0.65735 | 0.66585 | 0.66696 | 0.66712 | -0.00932 | -0.00082 | 0.00078 | 0.00045 | 0.00083 | 0.00073 | 0.00048 | 0.00048 |
| 400 | 0.65893 | 0.66595 | 0.66739 | 0.66750 | -0.00773 | -0.00072 | 0.00072 | 0.00083 | 0.00060 | 0.00049 | 0.00037 | 0.00037 |
| 500 | 0.66032 | 0.66638 | 0.66733 | 0.66740 | -0.00635 | -0.00029 | 0.00066 | 0.00073 | 0.00051 | 0.00042 | 0.00030 | 0.00030 |
| 750 | 0.66167 | 0.66666 | 0.66674 | 0.66678 | -0.00500 | -0.00027 | 0.00045 | 0.00011 | 0.00037 | 0.00028 | 0.00020 | 0.00020 |
| 1000 | 0.66274 | 0.66693 | 0.66690 | 0.66692 | -0.00392 | 0.00026 | 0.00024 | 0.00026 | 0.00028 | 0.00021 | 0.00015 | 0.00015 |

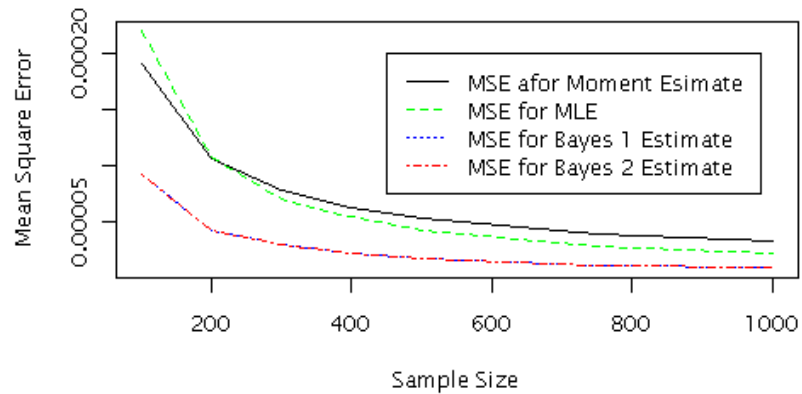

Fig.-1: Plots of trend in MSE of θ for varying n at θ=0.1

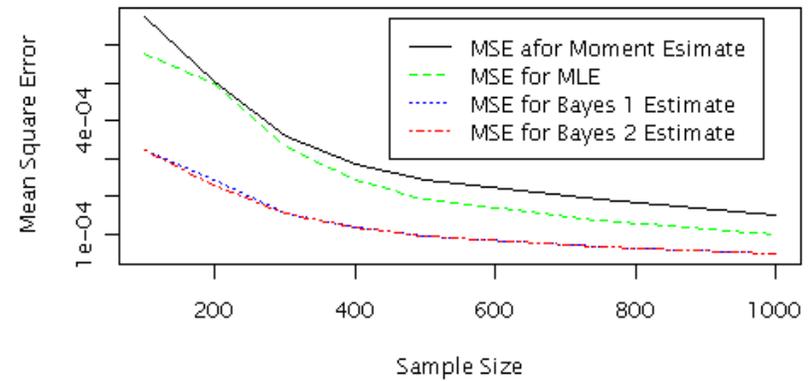

Fig.-2: Plots of trend in MSE of θ for varying n at θ=0.25

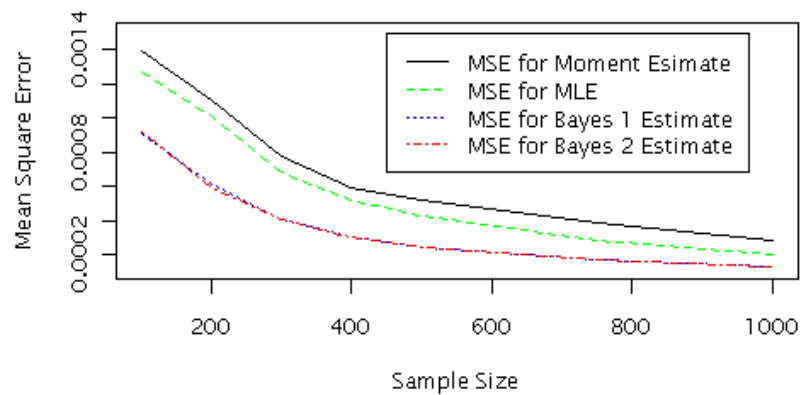

Fig.-3: Plots of trend in MSE of θ for varying n at θ=0.5

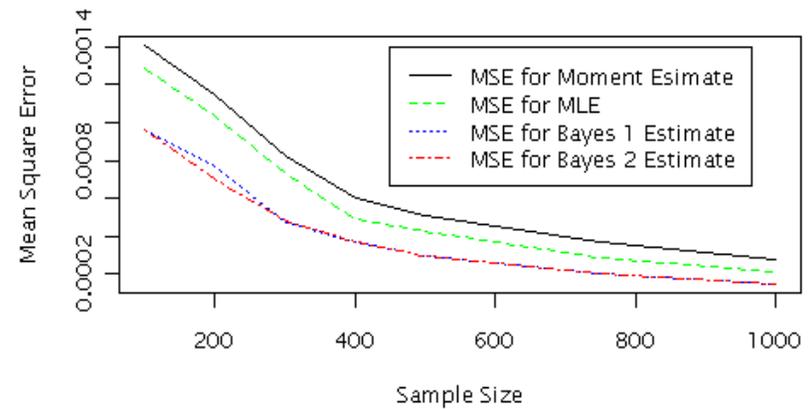

Fig.-4: Plots of trend in MSE of θ for varying n at θ=0.67